\documentstyle{camera}

\newcommand{\AmS}{{\protect\the\textfont2
  A\kern-.1667em\lower.5ex\hbox{M}\kern-.125emS}}

\newcommand{\beq}{\begin{equation}}
\newcommand{\eeq}{\end{equation}}
\newcommand{\bea}{\begin{eqnarray}}
\newcommand{\eea}{\end{eqnarray}}

\def\dm2{\Delta m^2}
\def\sq2{sin^2(2\Theta)}

\input epsf

\begin{document}

\title{DARK ENERGY MODELS TOWARD OBSERVATIONAL TESTS AND DATA}

\author{SALVATORE CAPOZZIELLO}

%
\organization{Dipartimento di Scienze Fisiche, Universit\`a di
Napoli "Federico II" and INFN, Sez. di Napoli, Italy}

\maketitle

\begin{abstract}
A huge amount of  good quality  data converges towards the picture
of a spatially flat universe undergoing the today observed phase
of accelerated expansion. This new observational trend is commonly
addressed as {\it Precision Cosmology}. Despite of the excellent
surveys, the nature of  dark energy, dominating the matter-energy
content of the universe, is still unknown and a lot of different
scenarios are viable candidates to explain cosmic acceleration.
Methods  to test these cosmological models are based on distance
measurements and lookback time toward astronomical objects used as
standard candles. The related degeneracy problem is the signal
that more data at low $(z\sim 0\div 1)$, medium $(1<z<10)$ and
high $(10 <z< 1000)$ redshift are needed to definitively select
realistic models.
\end{abstract}
\vspace{1.0cm}

\section{Introduction}

The increasing bulk of data that have been accumulated  in the
last few years have paved the way to the emergence of a new
standard cosmological model usually referred to as the {\it
concordance model}. The Hubble diagram of Type Ia Supernovae
(hereafter SNeIa) has been the first evidence  that the universe
is undergoing a phase of accelerated expansion. On the other hand,
balloon born experiments determined the location of the first and
second peak in the anisotropy spectrum of  cosmic microwave
background radiation (CMBR) strongly pointing out that the
geometry of the universe is spatially flat. If combined with
constraints coming from galaxy clusters on the matter density
parameter $\Omega_M$, these data indicate that the universe is
dominated by a non-clustered fluid with negative pressure,
generically dubbed {\it dark energy}, which is able to drive the
accelerated expansion. This picture has been further strengthened
by  more precise measurements of the CMBR spectrum, due to the
WMAP experiment \cite{WMAP}, and by the extension of the SNeIa
Hubble diagram to redshifts higher than 1.

After these observational evidences, an  overwhelming flood of
papers, presenting a great variety of models trying to explain
this phenomenon, has appeared; in any case,  the simplest
explanation is claiming for the well known cosmological constant
$\Lambda$. Although the best fit to most of the available data
\cite{WMAP}, the $\Lambda$CDM model failed in explaining why the
inferred value of $\Lambda$ is so tiny (120 orders of magnitude
lower) compared to the typical vacuum energy values predicted by
particle physics and why its energy density is today comparable to
the matter density  (the so called {\it coincidence problem}). As
a tentative solution, many authors have replaced the cosmological
constant with a scalar field rolling down its potential and giving
rise to the class of models now referred to as {\it quintessence}.
Even if successful in fitting the data, the quintessence approach
to dark energy is still plagued by the coincidence problem since
the dark energy and matter densities evolve differently and reach
comparable values for a very limited portion of the universe
evolution,  coinciding at present era. In this case, the
coincidence problem is replaced with a fine-tuning problem.
Moreover, it is not clear where this scalar field originates from,
thus leaving a great uncertainty on the choice of the scalar field
potential.

The subtle and elusive nature of  dark energy has led many authors
to  look for completely different scenarios able to give a
quintessential behavior without the need of exotic components. To
this aim, it is worth stressing that the acceleration of the
universe only claims for a negative pressure dominant component,
but does not tell anything about the nature and the number of
cosmic fluids filling the universe. This consideration suggests
that it could be possible to explain the accelerated expansion by
introducing a single cosmic fluid with an equation of state
causing it to act like dark matter at high densities and dark
energy at low densities. An attractive feature of these models,
usually referred to as {\it Unified Dark Energy} (UDE) or {\it
Unified Dark Matter} (UDM) models, is that such an approach
naturally solves, al least phenomenologically, the coincidence
problem. Some interesting examples are the generalized Chaplygin
gas and the tachyon fields. A different class of UDE models has
been proposed \cite{Hobbit} where a single fluid is considered
whose energy density scales with the redshift in such a way that
the radiation dominated era, the matter dominated era and the
accelerating phase can be naturally achieved. It is worth noting
that such class of models are extremely versatile since they can
be interpreted both in the framework of UDE models and as a
two-fluid scenario with dark matter and scalar field dark energy.
The main ingredient of the approach is that a generalized equation
of state can be always obtained and observational data can be
fitted.

Actually, there is still a different way to face the problem of
cosmic acceleration. It is possible that the observed acceleration
is not the manifestation of another ingredient in the cosmic pie,
but rather the first signal of a breakdown of our understanding of
the laws of gravitation. From this point of view, it is thus
tempting to modify the Friedmann equations to see whether it is
possible to fit the astrophysical data with  models comprising
only the standard matter. An interesting example of this kind is
the DGP gravity \cite{DGP}.

Moving in this framework, it is possible to  find alternative
schemes where a quintessential behavior is obtained by taking into
account effective models coming from fundamental physics giving
rise to generalized or higher order gravity actions (see for
example \cite{curvature}). SNeIa data could also be efficiently
fitted including higher-order curvature invariants in the gravity
Lagrangian.

It is worth noticing that these alternative schemes provide
naturally a cosmological component with negative pressure whose
origin is  related to the geometry of the universe thus overcoming
the problems linked to the physical significance of the scalar
field.

It is evident, from this short overview, the high number of
cosmological models which are viable candidates to explain the
observed accelerated expansion. This abundance of models is from
one hand the signal of the fact that we have a limited number of
cosmological tests to discriminate among rival theories, and from
the other hand that a urgent degeneracy problem has to be faced.
To this aim, it is useful to remember that both the SNeIa Hubble
diagram and the angular size\,-\,redshift relation of compact
radio sources  are distance based methods to probe cosmological
models so then systematic errors and biases could be iterated.
From this point of view, it is interesting to look for tests based
on time-dependent observables. For example, one can take into
account the {\it lookback time} to distant objects since this
quantity can discriminate among different cosmological models.

The lookback time is observationally estimated as the difference
between the present day age of the universe and the age of a given
object at redshift $z$. Such an estimate is possible if the object
is a galaxy observed in more than one photometric band since its
color is determined by its age as a consequence of stellar
evolution. It is thus possible to get an estimate of the galaxy
age by measuring its magnitude in different bands and then using
stellar evolutionary codes to choose the model that reproduces the
observed colors at best. A  similar approach was pursued by Lima
\& Alcaniz \cite{LA00} who used the age (rather than the lookback
time) of old high redshift galaxies to constrain the dark energy
equation of state. It is worth noting, however, that the estimate
of the  age of a single galaxy may be affected by systematic
errors which are difficult to control. Actually, this problem can
be overcome by considering a sample of galaxies belonging to the
same cluster. In this way, by averaging the estimates of all
galaxies, one obtains an estimate of the cluster age and reduces
the systematic errors. Such a method was first proposed in
\cite{DAJM01} and then used in \cite{FMT03} to test a class of
models where a scalar field is coupled with the matter term giving
rise to a particular quintessence model.

In this report, I shortly  discuss the dark energy "paradigm" and
some methods to constrain it toward observational data. Far from
being exhaustive and complete, my aim is to point out the
degeneracy problem and the fact that we need further and
self-consistent observational surveys at {\it all} redshifts to
remove it.

\section{The dark energy "paradigm"}

Many rival theories  have been advocated to fit the observed
accelerated expansion and to solve the puzzle  of the dark energy.
As a simple classification scheme, we may divide the different
cosmological models in three wide classes. According to the models
of the first class, the dark energy is a new ingredient of the
cosmic Hubble flow, the simplest case being the $\Lambda$CDM
scenario and its quintessential generalization which we will refer
to as QCDM models.

This is in sharp contrast with the assumption of UDE models (the
second class) where there is a single fluid described by an
equation of state comprehensive of all regimes of cosmic evolution
\cite{Hobbit} which I will consider here referring to it as the
{\it parametric density models} or generalized Equation of State
{\it EoS} models.

Finally, according to the third class models, accelerated
expansion could be the first evidence of a breakdown of the
Einstein General Relativity (and thus of the Friedmann equations)
which has to be considered as a particular case of a more general
theory of gravity. As an example of this kind of models, one can
consider the $f(R)$\,-\,gravity \cite{curvature}.

Far from being exhaustive, considering these three classes of
models allows to explore very different scenarios proposed to
explain the observed cosmic acceleration. However, the "paradigm"
is the $\Lambda$CDM model.

Cosmological constant $\Lambda$ has  become  a textbook candidate
to drive the accelerated expansion of the spatially flat universe.
Despite its {\it conceptual} problems, the $\Lambda$CDM model
turns out to be the best fit to a combined analysis of completely
different astrophysical data ranging from SNeIa to CMBR anisotropy
spectrum and galaxy clustering \cite{WMAP,SDSS03}.

As a simple generalization, one may consider the QCDM scenario in
which the barotropic factor $w \equiv p/\rho$ takes at a certain
epoch a negative value with $w = -1$ corresponding to the standard
cosmological constant. Testing whether  such a barotropic factor
deviate or not from $-1$ is one of the main issue of modern
observational cosmology. How such a negative pressure fluid drives
the cosmic acceleration may be easily understood by looking at the
Friedmann equations\,:

\begin{equation}
H^2 \equiv \left ( \frac{\dot{a}}{a} \right )^2 = \frac{8 \pi
G}{3} (\rho_{M} + \rho_Q) \ , \label{eq: fried1}
\end{equation}

\begin{equation}
2 \frac{\ddot{a}}{a} + H^2 = - 8 \pi G p_Q = - 8 \pi G w \rho_Q \
, \label{eq: fried2}
\end{equation}
where the dot denotes differentiation with respect to  cosmic time
$t$, $H$ is the Hubble parameter and the universe is assumed
spatially flat as suggested by the position of the first peak in
the CMBR anisotropy spectrum.

From the continuity equation, $\dot{\rho} + 3 H (\rho + p) = 0$,
we get for the $i$\,-\,th fluid with $p_i = w_i \rho_i$\,:

\begin{equation}
\Omega_i = \Omega_{i,0} a^{-3 (1 + w_i)} = \Omega_{i,0} (1 + z)^{3
(1 + w_i)} \ , \label{eq: omegavsz}
\end{equation}
where $z \equiv 1/a - 1$ is the redshift, $\Omega_i =
\rho_i/\rho_{crit}$  is the density parameter for the $i$\,-\,th
fluid in terms of the critical density $\rho_{crit} = 3H_0^2/8\pi
G$ and, hereafter, I label all the quantities evaluated today with
a subscript $0$. Inserting this result into Eq.(\ref{eq: fried1}),
one gets\,:

\begin{equation}
H(z) = H_0 \sqrt{\Omega_{M,0} (1 + z)^3 + \Omega_{Q,0} (1 + z)^{3
(1 + w)}} \ . \label{eq: hvsz}
\end{equation}
Using Eqs.(\ref{eq: fried1}), (\ref{eq: fried2}) and the
definition of the deceleration parameter $q \equiv - a
\ddot{a}/\dot{a}^2$, one finds\,:

\begin{equation}
q_0 = \frac{1}{2} + \frac{3}{2} w (1 - \Omega_{M,0}) \ .
\label{eq: qlambda}
\end{equation}
The SNeIa Hubble diagram, the large scale galaxy clustering  and
the CMBR anisotropy spectrum can all be fitted by the $\Lambda$CDM
model with $(\Omega_{M,0}, \Omega_Q) \simeq (0.3, 0.7)$ thus
giving $q_0 \simeq -0.55$, i.e. the universe turns out to be in an
accelerated expansion phase. The simplicity of the model and its
capability of fitting the most of the data are the reasons why the
$\Lambda$CDM scenario is the leading candidate to explain the dark
energy cosmology. Nonetheless,  its generalization,  QCDM models,
i.e.  mechanisms allowing the evolution of $\Lambda$ from the past
are invoked  to remove the $\Lambda$-problem and the $coincidence$
problem.

\section{Methods to constrain models}
Let us discuss now how cosmological models can be constrained
using suitable distance and/or time indicators. As a general
remark,  solutions coming from cosmological models have to be
matched with observations by using the redshift $z$ as the natural
time variable for the Hubble parameter, i.e.

\begin{equation} H(z)=-\frac{\dot{z}}{z+1}\,. \end{equation}  Interesting
ranges for $z$  are: $100< z < 1000$ for early universe (CMBR
data), $10 < z < 100$ (LSS), $0 < z < 10$ (SNeIa, radio-galaxies,
GRBs, etc.). The method consists in building up a reasonable
patchwork of data coming from different epochs and then matching
them with the same cosmological solution ranging, in principle,
from inflation to present accelerated era.

In order to constrain the parameters characterizing the
cosmological solution, a reasonable approach is to maximize the
following likelihood function\,:

\begin{equation}
{\cal{L}} \propto \exp{\left [ - \frac{\chi^2({\bf p})}{2} \right
]} \label{eq: deflike1}
\end{equation}
where {\bf p} are the parameters assigning  the cosmological
solution. The $\chi^2$ merit function can be defined as\,:

\begin{equation}
\chi^2({\bf p})  =  \sum_{i = 1}^{N}{\left [ \frac{y^{th}(z_i,
{\bf p}) - y_i^{obs}}{\sigma_i} \right ]^2}  + \displaystyle{\left
[ \frac{{\cal{R}}({\bf p}) - 1.716}{0.062} \right ]^2} +
\displaystyle{\left [ \frac{{\cal{A}}({\bf p}) - 0.469}{0.017}
\right ]^2}  \ . \label{eq: defchi1}\
\end{equation}

Terms entering Eq.(\ref{eq: defchi1}) can be characterized as
follows. For example, the dimensionless coordinate distances $y$
to objects at redshifts $z$ are considered in the first term. They
are defined as\,:

\begin{equation}
y(z) = \int_{0}^{z}{\frac{dz'}{E(z')}} \label{eq: defy}
\end{equation}
where $E(z)=H(z)/H_0$ is the normalized Hubble parameter. This is
the main quantity which allows to compare the theoretical results
with data. The function $y$ is related to the luminosity distance
$D_L = (1 + z) r(z)$.  A sample of data on $y(z)$ for the 157
SNeIa is discussed in the Riess et al. \cite{Riess04} Gold dataset
and 20 radio-galaxies are in \cite{DD04}. These authors fit with
good accuracy the linear Hubble law at low redshift ($z < 0.1$)
obtaining the Hubble dimensionless parameter $h = 0.664 {\pm}
0.008\,.$  Such a number can be consistently taken into account at
low redshift.  This value  is in agreement with $H_0 = 72 {\pm} 8
\ {\rm km \ s^{-1} \ Mpc^{-1}}$ given by the HST Key project
\cite{Freedman} based on the local distance ladder and estimates
coming from  time delays in multiply imaged quasars  and
Sunyaev\,-\,Zel'dovich effect in X\,-\,ray emitting clusters
\cite{H0SZ}.

The second term in Eq.(\ref{eq: defchi1}) allows to extend the
$z$-range to probe $y(z)$ up to the last scattering surface
$(z\geq 1000)$.  The {\it shift parameter} \cite{WM04,WT04} $
{\cal R} \equiv \sqrt{\Omega_M} y(z_{ls}) $ can be determined from
the CMBR anisotropy spectrum, where $z_{ls}$ is the redshift of
the last scattering surface which can be approximated as  $ z_{ls}
= 1048 \left ( 1 + 0.00124 \omega_b^{-0.738} \right ) \left ( 1 +
g_1 \omega_M^{g_2} \right ) $ with $\omega_i = \Omega_i h^2$ (with
$i = b, M$ for baryons and total matter respectively) and $(g_1,
g_2)$ given in \cite{HS96}. The parameter $\omega_b$ is
constrained by the baryogenesis calculations contrasted to the
observed abundances of primordial elements. Using this method, the
value $ \omega_b = 0.0214 {\pm} 0.0020  $  is found \cite{Kirk}.
In any case, it is worth noting that the exact value of $z_{ls}$
has a negligible impact on the results and setting $z_{ls} = 1100$
does not change constraints and priors on the other  parameters of
the given model. The third term in the function $\chi^2$ takes
into account  the {\it acoustic peak} of the large scale
correlation function at $100 \ h^{-1} \ {\rm Mpc}$ separation,
detected by using  46748 luminous red galaxies (LRG) selected from
the SDSS Main Sample \cite{Eis05}. The quantity

\begin{equation}
{\cal{A}} = \frac{\sqrt{\Omega_M}}{z_{LRG}} \left [
\frac{z_{LRG}}{E(z_{LRG})} y^2(z_{LRG}) \right ]^{1/3} \label{eq:
defapar}
\end{equation}
is related to the position of acoustic peak where $z_{LRG} = 0.35$
is the effective redshift of the above sample. The parameter
${\cal{A}}$ depends  on the dimensionless coordinate distance (and
thus on the integrated expansion rate),  on $\Omega_M$ and $E(z)$.
This dependence removes some of the degeneracies intrinsic in
distance fitting methods. Due to this reason, it is particularly
interesting to include ${\cal{A}}$ as a further constraint on the
model parameters using its measured value  $ {\cal{A}} = 0.469
{\pm} 0.017  $  \cite{Eis05}. Note that, although similar to the
usual $\chi^2$ introduced in statistics, the reduced $\chi^2$
(i.e., the ratio between the $\chi^2$ and the number of degrees of
freedom) is not forced to be 1 for the best fit model because of
the presence of the priors on ${\cal{R}}$ and ${\cal{A}}$ and
since the uncertainties $\sigma_i$ are not Gaussian distributed,
but take care of both statistical errors and systematic
uncertainties. With the definition (\ref{eq: deflike1}) of the
likelihood function, the best fit model parameters are those that
maximize ${\cal{L}}({\bf p})$.

Using the method sketched above, the several classes of models can
be constrained and selected by  observations. However, most of the
tests recently used to constrain cosmological parameters (such as
the SNeIa Hubble diagram and the angular size\,-\,redshift) are
essentially distance\,-\,based methods. The proposal of Dalal et
al. \cite{DAJM01} to use the lookback time to high redshift
objects is thus particularly interesting since it relies on a
completely different observable. The lookback time is defined as
the difference between the present day age of the universe and its
age at redshift $z$ and may be computed as\,:

\begin{equation}
t_L(z, {\bf p}) = t_H \int_{0}^{z}{\frac{dz'}{(1 + z') E(z', {\bf
p})}} \label{eq: deftl}
\end{equation}
where $t_H = 1/H_0 = 9.78 h^{-1} \ {\rm Gyr}$ is the Hubble  time
(with $h$ the Hubble constant in units of $100 \ {\rm km \ s^{-1}
\ Mpc^{-1}}$),  and, as above, $E(z, {\bf p}) = H(z)/H_0$ is the
dimensionless Hubble parameter and  $\{{\bf p}\}$ the set of
parameters characterizing the cosmological model. It is worth
noting that, by definition, the lookback time is not sensible to
the present day age of the universe $t_0$ so that it is (at least
in principle) possible that a model fits well the data on the
lookback time, but nonetheless it predicts a completely wrong
value for $t_0$. This latter parameter can be evaluated from
Eq.(\ref{eq: deftl}) by simply changing the upper integration
limit from $z$ to infinity. This shows that it is indeed a
different quantity since it depends on the full evolution of the
universe and not only on how the universe evolves from the
redshift $z$ to now. That is why this quantity can be explicitly
introduced as a further constraint. As an example, let us discuss
how to use the lookback time and the age of the universe to test a
given cosmological model. To this end, let us consider an object
$i$ at redshift $z$ and denote by $t_i(z)$ its age defined as the
difference between the age of the universe when the object was
born, i.e. at the formation redshift $z_F$, and the one at $z$. It
is\,:

\begin{equation}
t_i(z)  = \int_{z}^{z_F}{\frac{dz'}{(1 + z') E(z', {\bf p})}}
 =  t_L(z_F) - t_L(z) \ . \label{eq: titl}
\end{equation}
where I have used the definition (\ref {eq: deftl}) of the
lookback time. Suppose now we have $N$ objects and we have been
able to estimate the age $t_i$ of the object at redshift $z_i$ for
$i = 1, 2, \ldots, N$. Using the previous relation, we can
estimate the lookback time $t_{L}^{obs}(z_i)$ as\,:

\begin{equation}
t_{L}^{obs}(z_i) = t_{0}^{obs} - t_i(z) - df \ , \label{eq:
deftlobs}
\end{equation}
where $t_{0}^{obs}$ is the today estimated age of  the universe
and a {\it delay factor} can be defined as $ df = t_0^{obs} -
t_L(z_F)\,. $ The delay factor is introduced to take into account
our ignorance of the formation redshift $z_F$ of the object.
Actually, what can be measured is the age $t_i(z)$ of the object
at redshift $z$. To estimate $z_F$, one should use Eq.(\ref{eq:
titl}) assuming a background cosmological model. Since our aim is
to determine what is the background cosmological model, it is
clear that we cannot infer $z_F$ from the measured age so that
this quantity is completely undetermined. It is worth stressing
that, in principle, $df$ should be different for each object in
the sample unless there is a theoretical reason to assume the same
redshift at the formation of all the objects. If this is indeed
the case, then it is computationally convenient to consider $df$
rather than $z_F$ as the unknown parameter to be determined from
the data. Again a likelihood function can be defined as\,:

\begin{equation}
{\cal{L}}_{lt}({\bf p}, df) \propto \exp{[-\chi^{2}_{lt}({\bf p},
df)/2]} \label{eq: deflikelt}
\end{equation}
with\,:

\begin{equation}
\chi^{2}_{lt} = \displaystyle{\frac{1}{N - N_p + 1}} \left \{
\left [ \frac{t_{0}^{theor}({\bf p}) -
t_{0}^{obs}}{\sigma_{t_{0}^{obs}}} \right ]^2
 + \sum_{i = 1}^{N}{\left [ \frac{t_{L}^{theor}(z_i, {\bf p}) -
t_{L}^{obs}(z_i)}{\sqrt{\sigma_{i}^2 + \sigma_{t}^{2}}} \right
]^2} \right \} \label{eq: defchi}
\end{equation}
where $N_p$ is the number of parameters of the model,  $\sigma_t$
is the uncertainty on $t_{0}^{obs}$, $\sigma_i$ the one on
$t_{L}^{obs}(z_i)$ and the superscript {\it theor} denotes the
predicted values of a given quantity. Note that the delay factor
enters the definition of $\chi^2_{lt}$ since it determines
$t_{L}^{obs}(z_i)$ from $t_i(z)$ in virtue of Eq.(\ref{eq:
deftlobs}), but the theoretical lookback time does not depend on
$df$. In principle, such a method should work efficiently to
discriminate among the various dark energy models. Actually, this
is not exactly the case due to the paucity of the available data
which leads to large uncertainties on the estimated parameters. In
order to partially alleviate this problem, it is convenient to add
further constraints on the models by using  Gaussian
priors\footnote{The need for  priors to reduce the parameter
uncertainties is often advocated for  cosmological tests. For
instance, in \cite{LA00} a strong prior on $\Omega_M$ is
introduced to  constrain the dark energy equation of state. It is
likely, that extending the dataset to higher redshifts and
reducing the uncertainties on the age estimate will allow to avoid
resorting to priors.} on the Hubble constant, i.e. redefining the
likelihood function as\,:

\begin{equation}
{\cal{L}}({\bf p}) \propto {\cal{L}}_{lt}({\bf p}) \exp{\left [
-\frac{1}{2} \left ( \frac{h - h^{obs}}{\sigma_h}  \right )^2
\right ]} \propto \exp{[- \chi^2({\bf p})/2]} \label{eq: deflike}
\end{equation}
where we have absorbed $df$ in the set of parameters ${\bf p}$ and
have defined\,:
\begin{equation}
\chi^2 = \chi^{2}_{lt} + \left ( \frac{h - h^{obs}}{\sigma_h}
\right )^2 \label{eq: newchi}
\end{equation}
with $h^{obs}$ the estimated value of $h$ and  $\sigma_h$ its
uncertainty. The HST Key project results \cite{Freedman} can be
used setting $(h, \sigma_h) = (0.72, 0.08)$. Note that this
estimate is independent of the cosmological model since it has
been obtained from local distance ladder methods. The best fit
model parameters ${\bf p}$ may  be obtained by maximizing
${\cal{L}}({\bf p})$ which is equivalent to minimize the $\chi^2$
defined in Eq.(\ref{eq: newchi}). It is worth stressing that such
a function should not be considered as a {\it statistical
$\chi^2$} in the sense that it is not forced to be of order 1 for
the best fit model to consider a fit as successful. Actually, such
an interpretation is not possible since the errors on the measured
quantities (both $t_i$ and $t_0$) are not Gaussian distributed
and, moreover, there are uncontrolled systematic uncertainties
that may also dominate the error budget. Nonetheless, a
qualitative comparison among different models may be obtained by
comparing the values of this pseudo $\chi^2$ even if this should
not be considered as a definitive evidence against a given model.
Having more than one parameter, one obtains the best fit  value of
each single parameter $p_i$ as the value which maximizes the
marginalized likelihood for that parameter defined as\,:

\begin{equation}
{\cal{L}}_{p_i} \propto \int{dp_1 \ldots \int{dp_{i - 1}
\int{dp_{i + 1} \ldots \int{dp_n \ {\cal{L}}({\bf p})}}}} \ .
\label{eq: deflikemar}
\end{equation}
After having normalized  the marginalized likelihood to 1  at
maximum, one computes the $1 \sigma$ and $2 \sigma$ confidence
limits  on that parameter by solving ${\cal{L}}_{p_i} =
\exp{(-0.5)}$ and ${\cal{L}}_{p_i} = \exp{(-2)}$ respectively. In
summary, taking into account the above procedures for distance and
time measurements, one can reasonably constrain a given
cosmological model. In any case, the main and obvious issue is to
have at disposal sufficient and good quality data sets.

\section{Conclusions}

The impressive amount of data   indicating a spatially flat
universe in accelerated expansion has posed the problem of dark
energy and stimulated the search for cosmological models able to
explain such unexpected behavior. Several theories have been
proposed to solve the puzzle of the nature of  dark energy ranging
from a rolling scalar field to a unified picture where a single
exotic fluid accounts for the whole dark sector (dark matter and
dark energy). Moreover, modifications of the gravity action has
also been advocated. Although deeply different in their underlying
physics, all these scenarios share the common feature of well
reproducing the available astrophysical data giving rise to a
degeneracy problem.  It is worth stressing, however, that the most
widely used cosmological tests (in particular the SNeIa Hubble
diagram and the angular size\,-\,redhisft relation) are
essentially based on distance measurements to high redshift
objects and are thus affected by similar systematic errors. It is
hence particularly interesting to look for  methods which are
related to the estimates of different quantities. Being affected
by other kinds of observational problems, such  methods could be
considered as cross checks for the results obtained by the usual
tests and they should represent complementary probes for
cosmological models. Among these alternative methods,  the
lookback time is related to a different astrophysics than the
distance based methods and it is thus free from any problem
connected with the evolution of standard candles (such as the
SNeIa absolute magnitude and the intrinsic linear size of radio
sources \cite{dunsby}). In any case, from one hand we need some
{\it experimentum crucis} capable of removing the degeneracy for a
reasonably large redshift range and, from the theoretical
viewpoint, we need a physically reliable cosmological model,
emerging from some fundamental theory, without the conceptual
shortcomings of $\Lambda$CDM.


\begin{thebibliography}{99}

\bibitem{WMAP}
 Spergel, D.N. et al.: 2003,  {\it Astroph. J. Supp.} {\bf 148},
 175.
\bibitem{Riess04}
 Riess A.G. et al.,: 2004,  {\it Astrophys.
J.} {\bf  607}, 665.
\bibitem{Hobbit}
  Capozziello S. et al.: 2006,   {\it Phys. Rev.} {\bf D 73},
  043512.
\bibitem{DGP}
 Dvali, G.R.,   Gabadadze G.,  Porrati,  M.,: 2000, {\it Phys. Lett.} {\bf B 485},
208.
\bibitem{curvature}
 Capozziello S.: 2002, {\it Int. J. Mod. Phys.} {\bf D 11}, 483.
\bibitem{LA00}
Lima, J.A.S.,  Alcaniz, J.S.: 2000,  {\it Mon. Not. Roy. Astr.
Soc.} {\bf 317}, 893.
\bibitem{DAJM01}
 Dalal N. et al.: 2001, {\it Phys. Rev. Lett.} {\bf 87}, 141302.
\bibitem{FMT03}
Ferreras, I. et al.: 2003, {\it Mon. Not. Roy. Astr. Soc.} {\bf
344}, 257.
\bibitem{SDSS03}
Tegmark, M. et al.: 2004,  {\it Phys. Rev.} {\bf D 69}, 103501.
\bibitem{Freedman}
Freedman, W.L. et al.: 2001,  {\it Astrophys. J.} {\bf  553}, 47.
\bibitem{DD04}
Daly, R.A., Djorgovski, S.G.,: 2004, astro\,-\,ph/0403664.
\bibitem{JDA03}
Jain, B., Dev, A.,  Alcaniz, J.S.,: 2003,  {\it Class. Quant.
Grav.} {\bf 20}, 4163.
\bibitem{H0SZ}
 Saunders, R. et al.,: 2003,  {\it Mon. Not. Roy. Astr.
Soc.} {\bf 341}, 937.
\bibitem{WM04}
 Wang, Y.,  Mukherjee, P., {\it Astroph. J}  {\bf 606}, 654.
\bibitem{WT04}
Wang, Y., Tegmark, M.,: 2004, {\it Phys. Rev. Lett.} {\bf 92},
241302.
\bibitem{HS96}
Hu, W.,  Sugiyama, N.,: 1996, {\it Astroph. J.} {\bf 471}, 542.
\bibitem{Kirk}
Kirkman, D. et al.,: 2003, {\it Astroph. J.} {\bf 149}, 1.
\bibitem{Eis05}
Eisenstein, D. et al.,: 2005, {\it Astroph. J.}  {\bf 633}, 560.
\bibitem{dunsby}
 Capozziello, S. et al.,: 2007,  {\it A$\&$A} {\bf 472},  51.

\end{thebibliography}
\end{document}